\pdfoutput=1
\documentclass[fleqn,usenatbib]{mnras}

\usepackage[T1]{fontenc}
\usepackage{ae,aecompl}

\usepackage{amssymb,amsmath}
\usepackage{graphicx}
\usepackage{aas_macros}
\usepackage[usenames]{color}

\newcommand\msun {M$_{\odot}$}

\def\lum{erg s$^{-1}$}

\def\chan{{\it Chandra}}
\def\ulxc{J120922+295551}
\def\ulxd{J120922+295559}
\def\ulxa{J022721+333500}
\def\ulxb{J022727+333443}
\def\ho2{Ho II X-1}
\def\h{{\it H}-band}

\def\vel{km s$^{-1}$}

\def\lx{$L_{\mathrm{X}}$}

\title[NIR spectroscopy of five ULX counterparts]{Keck/MOSFIRE spectroscopy of five ULX counterparts}
\author[M. Heida et al.]
{M. Heida$^{1,2,3}$, P. G. Jonker$^{2,3}$, M. A. P. Torres$^{2,4}$, T. P. Roberts$^5$, 
\newauthor D. J. Walton$^{6,1}$, D.-S. Moon$^7$, D. Stern$^6$, F. A. Harrison$^1$\\
$^1$Space Radiation Laboratory, California Institute of Technology, Pasadena, CA 91125, USA\\
$^2$SRON Netherlands Institute for Space Research, Sorbonnelaan 2, 3584 CA Utrecht, the Netherlands\\
$^3$Department of Astrophysics/IMAPP, Radboud University Nijmegen, P.O. Box 9010, 6500 GL Nijmegen, The Netherlands\\
$^4$European Southern Observatory, Alonso de Cordova 3107, Casilla 19001, Vitacura, Santiago 19, Chile\\
$^5$Department of Physics, University of Durham, South Road, Durham DH1 3LE, United Kingdom\\
$^6$Jet Propulsion Laboratory, California Institute of Technology, 4800 Oak Grove Drive, Pasadena, CA 91109, USA\\
$^7$Department of Astronomy and Astrophysics, University of Toronto, Toronto, ON M5S 3H4, Canada
}

\date{Accepted XXX. Received YYY; in original form ZZZ}

\pubyear{2016}

\begin{document}
\label{firstpage}
\pagerange{\pageref{firstpage}--\pageref{lastpage}}
\maketitle

% Abstract
\begin{abstract}
We present \h{} spectra of the candidate counterparts of five ULXs (two in NGC 925, two in NGC 4136, and Holmberg~II~X-1) obtained with Keck/MOSFIRE. The candidate counterparts of two ULXs (\ulxa{} in NGC 925 and \ulxd{} in NGC 4136) have spectra consistent with (M-type) red supergiants (RSGs). We obtained two epochs of spectroscopy of the candidate counterpart to \ulxa{}, separated by 10 months, but discovered no radial velocity variations with a 2-$\sigma$ upper limit of 40 \vel. If the RSG is the donor star of the ULX, the most likely options are that either the system is seen at low inclination ($< 40^\circ$), or the black hole mass is less than 100 \msun, unless the orbital period is longer than 6 years, in which case the obtained limit is not constraining.
The spectrum of the counterpart to \ulxd{} shows emission lines on top of its stellar spectrum, and the remaining three counterparts do not show absorption lines that can be associated with the atmosphere of a star; their spectra are instead dominated by emission lines.
Those counterparts with RSG spectra may be used in the future to search for radial velocity variations, and, if those are present, determine dynamical constraints on the mass of the accretor.
\end{abstract}

\begin{keywords}
Infrared: Stars -- X-rays: individual: Holmberg II X-1 -- X-rays: individual: [SST2011] J022721.52+333500.7 -- X-rays: individual: [SST2011] J022727.53+333443.0 -- X-rays: individual: CXOU J120922.6+295551 -- X-rays: individual: [SST2011] J120922.18+295559.7
\end{keywords}

\section{Introduction}
Ultraluminous X-ray sources (ULXs) are point-like, off-nuclear X-ray sources with a luminosity in the 0.3--10 keV band $> 10^{39}$ \lum{} (see \citealt{feng11} for a review). This is higher than the Eddington luminosity of a 10 \msun{} black hole (BH), which raises the question how ULXs attain their high luminosities. The majority of (candidate) ULXs, with \lx $< 10^{41}$ \lum, likely contain stellar mass BHs ($M_\mathrm{BH} \leq$ 20 \msun) accreting at or above their Eddington limit (\citealt{middleton15} and references therein), although some of these ULXs could still contain massive stellar ($M_\mathrm{BH} =$ 20 -- 100 \msun) or even intermediate mass BHs ($M_\mathrm{BH} =$ 100 -- 10$^5$ \msun; \citealt{colbert99}). The most luminous ULXs, with \lx $> 10^{41}$ \lum{} (referred to as hyperluminous X-ray sources or HLXs) are hard to explain as stellar mass BHs, and are strong candidates to host BHs more massive than those found in X-ray binaries in our Galaxy (\citealt{farrell09, strohmayer03, pasham14, casares14}).

To date, there are only two ULXs with mass estimates that do not depend on modelling of their X-ray spectra (\citealt{motch14,liu13}). In both cases, although the uncertainties on the masses are large, a stellar mass BH is favoured as the accretor. In addition, \citet{bachetti14} discovered X-ray pulsations from M82 X-2, proving that the accretor in that ULX is a neutron star.

Improved model-independent mass measurements of ULX accretors are necessary to prove that the so-called `ultraluminous state' (\citealt{gladstone09}) is indeed a sign of super-Eddington accretion onto stellar mass BHs, to see if neutron star accretors are common among ULXs, and to find out if some ULXs contain massive stellar or intermediate mass BHs. 
Dynamical mass measurements would be the most reliable method. Through phase-resolved spectroscopic observations, it may be possible to trace the radial velocity curve of the donor and measure the mass function, setting a lower limit on the mass of the compact object. In the optical regime, attempts to measure the mass function in this way have been hampered by the presence of bright accretion discs in these systems (\citealt{roberts11,liu12}). However, some ULXs may have red supergiant (RSG) donor stars and these can outshine the disc in the near-infrared (NIR) part of the spectrum (\citealt{copperwheat05}). 

In \citet{heida14} we describe our systematic search for NIR counterparts to nearby ULXs. We discovered 11 candidate counterparts that could be RSG donor stars. NIR spectroscopy with the Very Large Telescope (VLT) has shown that one of these candidate counterparts, to RX~J004722.4-252051 in NGC 253, is indeed an M-type RSG, with a peculiar velocity suggestive of a massive stellar BH accretor (\citealt{heida15a}). In this paper we describe the results of our observations with the Keck telescope that yielded NIR spectra of five other ULX counterparts from the \citet{heida14} sample. Two of the ULXs are situated in NGC 925 and two in NGC 4136. The fifth ULX is Holmberg II X-1 (hereafter \ho2). Three of these ULXs (\ulxa{} in NGC 925 and \ulxc{} and \ulxd{} in NGC 4136) have been observed to reach maximum X-ray luminosities of a few times $10^{39}$ \lum; \ulxb{} in NGC 925 and \ho2{} are brighter, reaching X-ray luminosities of a few times $10^{40}$ \lum{} (\citealt{swartz11, liu05, grise10}). 
The goal of this campaign was to characterize the NIR counterparts. Hence we initially obtained only one epoch of data per source. Based on these observations, we obtained a second epoch for one source in NGC 925, \ulxa, to search for radial velocity shifts. 

This paper is organised as follows: in Section 2 we describe the setup of our observations. Section 3 describes our data reduction and analysis. In Section 4 we present the results for the separate counterparts and in Section 5 we discuss our findings and conclude.

% Data description and reduction and analysis
\section{Observations}
%\subsection{Keck/MOSFIRE}
We obtained \h{} spectra of five ULX counterparts with the Multi-Object Spectrometer for Infra-Red Exploration (MOSFIRE; \citealt{mclean10,mclean12}), mounted on the Keck I telescope on Mauna Kea\footnote{The data will be publicly available 18 months after the observations in the Keck Observatory Archive (https://koa.ipac.caltech.edu)}. MOSFIRE has a field of view of $6.1' \times 6.1'$ with a pixel scale of $0.18''$ and a robotic slit mask system with up to 46 slits. 
The data were taken on 2013 December 22, 2014 January 10 (program ID C241M), and 2014 November 4 (program ID C201M; all dates are in UT). We used slit masks with $0.7''$ wide slits which give us, in combination with the \h{} filter and the fixed MOSFIRE diffraction grating, a spectral resolution $R \approx 3000$ and spectral coverage from $\sim14500$ -- 18000 \AA. The seeing on all nights was $> 0.7''$ ($\sim 0.8''$ -- $1.0''$), so that the resolution is set by the slit width. The integration time per exposure was 119.3 s on all nights. On 2013 December 22 and 2014 January 10 we used an ABBA nodding pattern with a nod amplitude of $1.5''$ along the slit. On 2014 November 4, we used a different slit orientation than we did for the observations of NGC 925 in December 2013 and January 2014 (see also Figure \ref{fig:ulxa_slits}). The orientation used in the first observations was chosen to avoid other nearby objects --- the new orientation was chosen to better separate the two components of the elongated counterpart of \ulxa; however, the seeing was not good enough to distinguish the two components. As a result there were multiple sources in the slit containing \ulxa{}, so we used an ABBA nodding pattern with a larger nod amplitude ($2.9''$ for the first 20 exposures and $3.4''$ for the remaining 71 exposures). The total exposure times are listed in Table \ref{tab:obs}. We observed telluric standard stars at similar airmass as the science targets before and after every series of exposures. No flux standards were observed. For every slit mask configuration we obtained flats and arc spectra in the afternoon before the observing run. 

\begin{table*}
  \centering
 \caption{Description of the Keck/MOSFIRE observations. Apparent and absolute \h{} magnitudes from \citet{heida14} are also provided. The last column describes the features that are visible in the NIR spectrum.}\label{tab:obs}
\begin{tabular}{lllllcl}
 \hline
 Host galacy & Source & \emph{H} & $M_H$ & Observation date (UT) & Time on source (h) & Spectral features\\
  \hline
 NGC 925 & \ulxa{} & 18.7 & $-10.6$ & 2013 Dec 22\& 2014 Jan 10 & 5.4 & RSG \\ % 82 x 120 s plus 80 x 120 s = 19440 s (really 119.29278 itime = 19325.4 s)
 & (Source 2) & & & 2014 Nov 4 & 3\\ % 91 x 120 s = 10920 s (really 10855.6 s)
NGC 925 &  \ulxb & 20.1 & $-9.2$ & 2013 Dec 22 \& 2014 Jan 10 & 5.4 & Em. lines\\
NGC 4136 & \ulxc & 19.1 & $-10.8$ & 2014 Jan 10 & 2.4 & Em. lines\\ % 72 x 120 s = 8640 s (really 8589.1 s)
NGC 4136 & \ulxd & 19.2 & $-10.7$ & 2014 Jan 10 & 2.4 & RSG + em. lines \\
Holmberg II & \ho2 & 20.6 & $-7.1$ & 2014 Jan 10 & 2.6 & Em. lines\\ % 78 x 120 = 9360 s (really 9304.8 s)
 \hline
 \end{tabular}
\end{table*}

\section{Data reduction and analysis}
We reduce the data with the MOSFIRE Data Reduction Pipeline version 2014.03.14 ({\sc MOSFIRE-DRP}; described in detail by \citealt{steidel14}). The pipeline handles the background subtraction and combines the separate exposures to produce a 2D, rectified and wavelength calibrated spectrum for every slit. The wavelength calibration is done by fitting a polynomial to the night sky lines, using wavelengths in vacuum. The typical standard deviation of the residuals after fitting is 0.1 \AA{} ($\sim$ 2 \vel{} at 1.65 $\mu$m). For every observing night and target, we split the science exposures into two groups of at most 1.5 hours and reduce them separately.

We then use the {\sc Starlink} program {\sc Figaro} to extract the spectra with the \emph{profile} and \emph{optextract} tasks (\citealt{horne86}). In the November 2014 observation of \ulxa{} we extract spectra of three objects that are visible in the slit. To correct for telluric absorption, we use the IDL routine {\sc Xtellcor\_general} (\citealt{vacca03}). 

For the final analysis (characterization of the spectra and --- if possible --- cross-correlation with template spectra to measure radial velocities) we use Tom Marsh's programme {\sc Molly}. We load the wavelength-calibrated spectra into {\sc Molly} and first run the task \emph{hfix} to calculate the heliocentric velocity. Because the spectra are not flux-calibrated, we normalize them by fitting a 3rd order polynomial and dividing by it. This removes any systematic flux offsets, which is important to do before we average the spectra of every target. The only exception is \ulxb, since it has no continuum that can be used for the normalization. Next, we rebin them to a common velocity scale of 30.25 \vel{} pixel$^{-1}$ using \emph{vbin}, which also moves the spectra to the heliocentric frame. Finally we average the spectra of every target per night. As expected, because the orbital periods for ULXs with an RSG donor are $\geq$ 3 years, we do not observe velocity shifts between the spectra taken in December 2013 and January 2014 and therefore we add them to increase the signal to noise ratio (S/N). 

For \ulxa{} and \ulxd, we use \emph{xcor} to cross-correlate the spectra with template spectra of a range of spectral types (that have been normalized in the same way as our science spectra). As templates we use model RSG spectra from \citet{lancon07}, with solar abundances, a mass of 15 \msun{} and temperatures ranging from 2900 -- 5000 K. In the spectra of \ulxa{} and \ulxd{} we mask the regions with high noise levels due to strong telluric emission lines. With \emph{xcor} we apply shifts to the spectrum of $-10$ to +30 pixels, with steps of 1 pixel, and calculate the value of the cross-correlation function for each pixel shift. \emph{xcor} then computes the radial velocity difference between spectrum and template by fitting a parabola to the cross-correlation function at the peak pixel and its two neighbouring pixels. 
We follow \citet{heida15a} to obtain a more robust measure of the uncertainty on the radial velocity measurements: we use the {\sc Molly} command \emph{boot} to produce 1000 bootstrapped copies of the spectra, and use \emph{xcor} to cross-correlate them with one of the model spectra (with T$_\mathrm{eff}$ = 3500 K). The resulting distribution of radial velocities is Gaussian. We fit a Gaussian curve to this distribution and adopt the centroid and width of this Gaussian as a robust estimate of the radial velocity and its uncertainty.
To check for signs of rotational broadening of the absorption lines in the RSG spectra, we broaden a template spectrum (with T$_\mathrm{eff}$ = 3500 K) with $10$ -- $200$ \vel{} in steps of $10$ \vel. We then cross-correlate those broadened template spectra with the RSG spectra to see if the cross-correlation improves for higher rotational velocities.

For the spectra where we detect emission lines, we use the {\sc Molly} command \emph{mgfit} to fit Gaussian profiles to the positions of [Fe {\sc ii}] and/or hydrogen Brackett lines in the spectra, and calculate their offset with respect to their rest wavelengths.

\section{Results}
\subsection*{NGC 925: \ulxa}
\begin{figure}
\begin{center}
\includegraphics[width=0.5\textwidth]{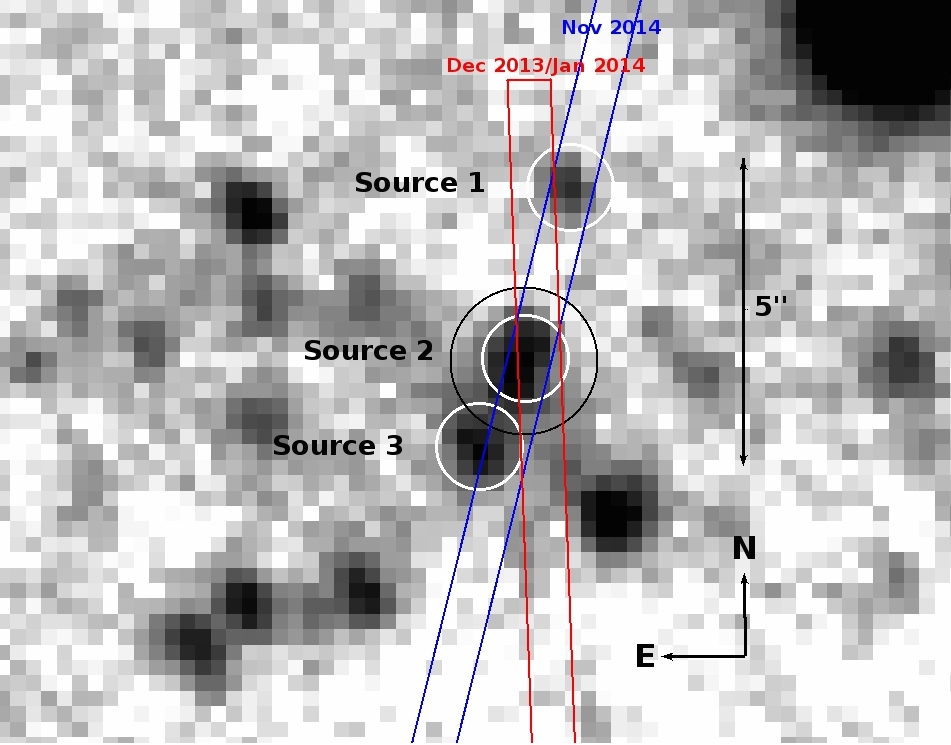}
\caption{WHT/LIRIS \h{} image of \ulxa{} in NGC 925, with the MOSFIRE slit positions indicated by red (December 2013/January 2014 observations) and blue (November 2014 observations) boxes. The black circle indicates the \chan{} X-ray localisation of the ULX. The white circles indicate the sources visible in our November 2014 MOSFIRE observation. Sources 3 is not centred on the slit, causing an offset $\leq 85 \pm 10$ \vel{} in its measured radial velocity.}\label{fig:ulxa_slits}
\end{center}
\end{figure}

\begin{figure*}
\includegraphics[width=\textwidth]{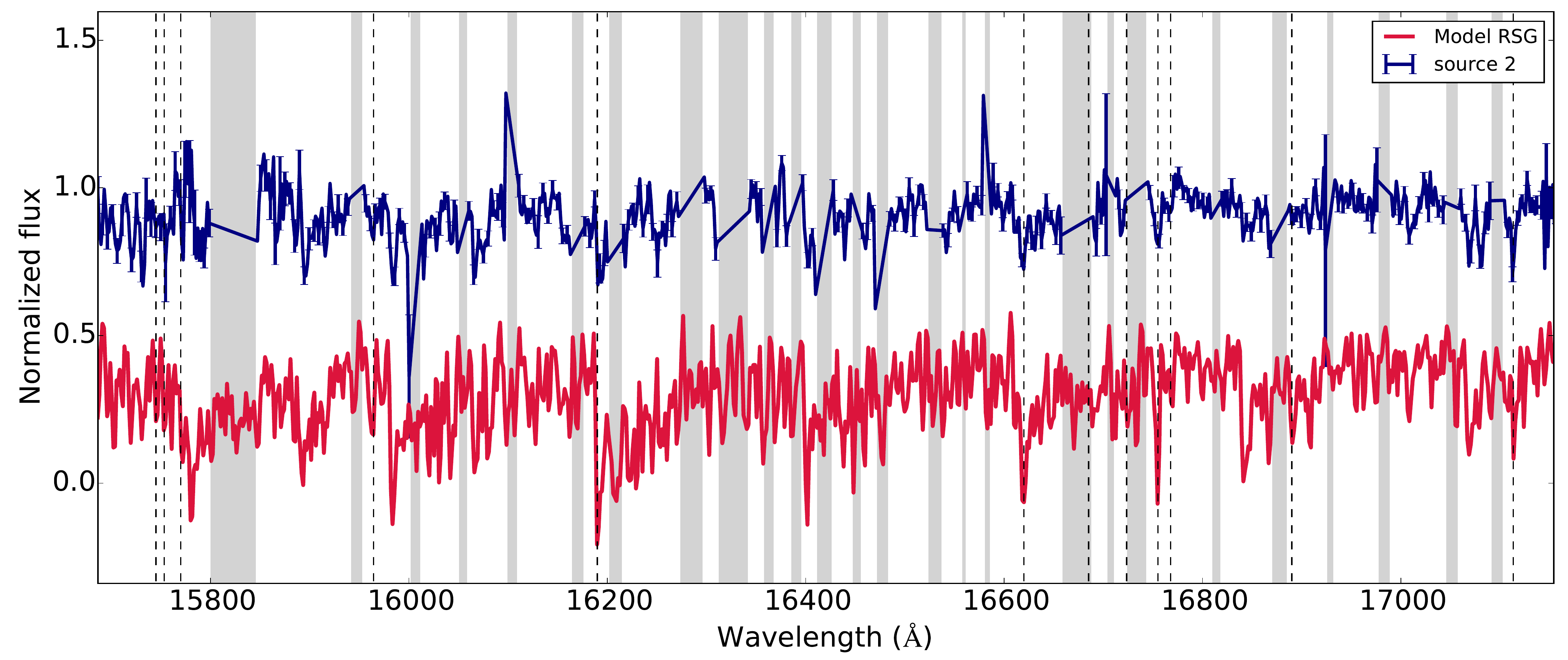}
\caption{The normalized spectra of NGC 925 \ulxa{} (source 2, from the Dec 2013/Jan 2014 observation, blue line) and a 3500 K model from \citet{lancon07} (red line, shifted downwards by 0.5 for clarity). The spectrum of \ulxa{} has been shifted by $-506$ \vel{} and has been interpolated over wavelength ranges that were strongly affected by noise from telluric emission lines (grey shaded areas). The `emission lines' visible near some of these ranges are due to this noise. The dashed lines indicate the positions of absorption lines that are indicative of late-type stars --- from short to long wavelength: Mg {\sc i} triplet (1.574, 1.575, 1.577 $\mu$m), Si {\sc i} (1.59 $\mu$m), CO bandhead (1.62 $\mu$m), CO bandhead (1.66 $\mu$m), Si {\sc i} (1.67 $\mu$m), Al {\sc i} triplet (1.672, 1.675, 1.676 $\mu$m), OH (1.69 $\mu$m), and Mg {\sc i} (1.71 $\mu$m).}\label{fig:ulxaspec}
\end{figure*}

In the NIR image of NGC 925 (see Figure \ref{fig:ulxa_slits}) the counterpart to \ulxa{} appears elongated. Due to the orientation of the slits that we used for the observations in December 2013 and January 2014, only one object (source 2 in Figure \ref{fig:ulxa_slits}) is visible in those observations. In the November 2014 observations we used a slit orientation to better separate the components of the counterpart, but the seeing was not good enough to resolve them. In these observations, spectra of three objects are visible. Source 2 (see Figure \ref{fig:ulxa_slits}) corresponds to the elongated counterpart. 
All three spectra (sources 1 to 3) show a continuum with many absorption lines from neutral metals. The CO bandheads at 1.62 and 1.66 $\mu$m are also detected, proving that the objects are all K- or M-type stars (see Figure \ref{fig:ulxaspec} for the spectrum of source 2). We find no evidence for velocity broadening of the absorption lines. This is fully consistent with what we expect for RSGs, which have rotational velocities of at most a few \vel, undetectable at MOSFIRE's resolution of $\sim80$ \vel{} in the \h.

From the cross-correlation with template spectra we find a radial velocity of $495 \pm 11$ \vel{} for source 1. For source 2 we find a radial velocity of $506 \pm 9$ \vel{} for the Dec 2013/Jan 2014 spectrum and $500 \pm 14$ \vel{} for the Nov 2014 spectrum. For source 3 we find a radial velocity of $502 \pm 8$ \vel{}. This source is not centred on the slit, which causes an offset in the radial velocity. The sources are too faint to be visible in our through-slit alignment images. However, brighter stars selected to check the slit alignment were detected and we use the comparison of their locations on the through slit image with that of the deep WHT/LIRIS image (\citealt{heida14}) to determine the actual location of the slit on the sky. From this we derive a best estimate of the offset between the location of source 3 and the centre of the slit of $0.5'' \pm 0.05''$ (see Figure \ref{fig:ulxa_slits}). This corresponds to $2.8 \pm 0.3$ pixels on the MOSFIRE detector, or a velocity offset of $85 \pm 10$ \vel{}. Because the seeing was comparable to the slit width, this value is an upper limit on the the velocity offset. If we correct for this we find a radial velocity $\geq 420 \pm 15$ \vel{} for source 3. The radial velocities of sources 1 and 2 are in agreement with what is expected from the radial velocity field of NGC 925 (\citealt{deblok08}). This proves that the objects are located in NGC 925, confirming the absolute magnitude reported for source 2 in \citet{heida14}. That absolute magnitude is such that even if source 2 consists of two equally bright sources, they are both RSGs. The radial velocity of source 3 is somewhat offset with respect to the local radial velocity, but that source is still most likely located in NGC 925.
The highest values for the cross-correlation function for all sources are found when cross-correlating with the templates with the lowest temperatures, indicating that these NIR sources are probably M-type stars. 

\subsection*{NGC 925: \ulxb}
\begin{figure*}
\hbox{
\includegraphics[width=\textwidth]{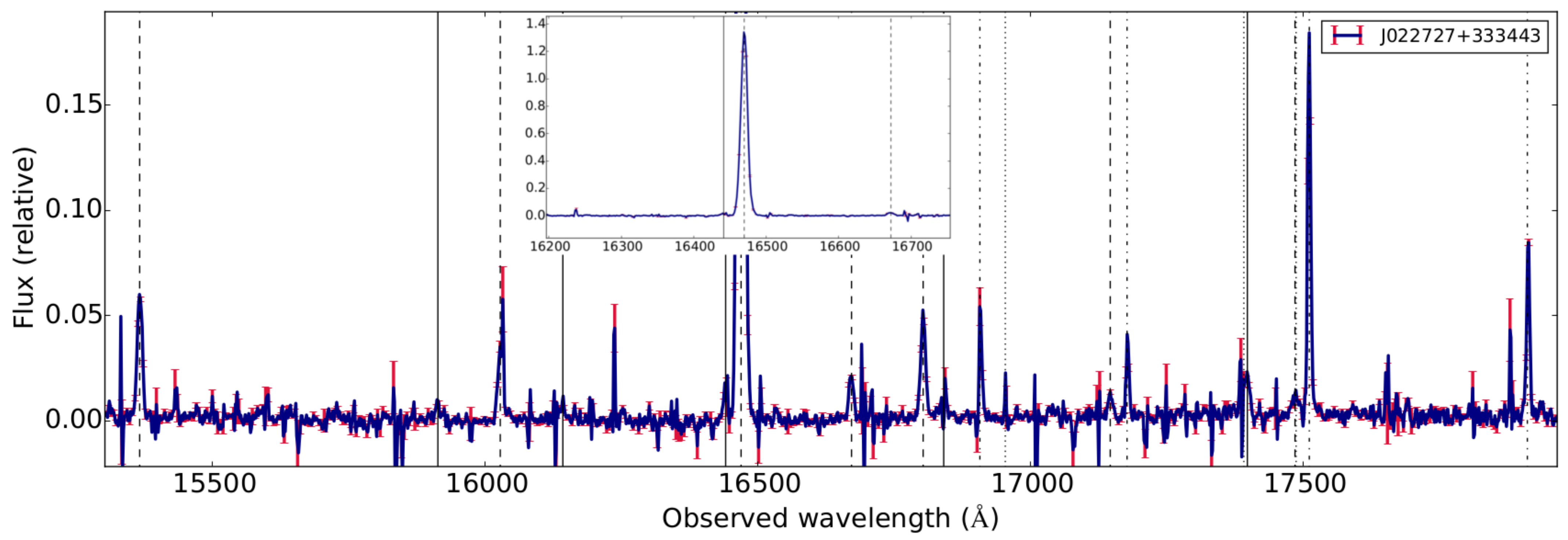}
}
\caption{The MOSFIRE spectrum of NGC 925 \ulxb. Error bars (in red) are plotted for every third data point. Dotted lines indicate positions of He lines, dashed lines indicate positions of [Fe {\sc ii}] lines, solid lines indicate the H Brackett series, dot-dashed lines indicate positions of H$_2$ lines. The vertical lines representing spectral transitions are redshifted by 540 \vel. The inset shows the full profile of the [Fe {\sc ii}]$\lambda$1.644 emission line.}\label{fig:ulxbspec}
\end{figure*}

The counterpart to \ulxb{} shows no continuum emission at all. The spectrum is dominated by a single strong emission line due to [Fe {\sc ii}] with a rest wavelength of $1.644 \mu$m. Many weaker emission lines are visible, due to [Fe {\sc ii}], H (Brackett lines), He I, He II and H$_2$. These are indicated in Figure \ref{fig:ulxbspec}. We fitted Gaussian profiles to these emission lines to calculate the radial velocity of the counterpart. The [Fe {\sc ii}], hydrogen and helium lines are well fitted by a Gaussian, but the H$_2$ lines are asymmetric. The radial velocity of the lines is $540 \pm 2$ \vel, proving that the source is part of NGC 925 and not a foreground or background object.
%No continuum, hence not normalised. Emission line spectrum. Strongest line Fe II 1644. 
%Lines from Fe II (fwhm 11 A), H brackett (fwhm 14 A), He I, II (fwhm 11 A), H$_2$ (fwhm 7 A). Verhoudingen H2 komen redelijk overeen met 4000 K. What does this mean, anyway? Vel offset $540 \pm 1$ \vel. 

%\pagebreak

\subsection*{\ho2}
\begin{figure*}
\includegraphics[width=\textwidth]{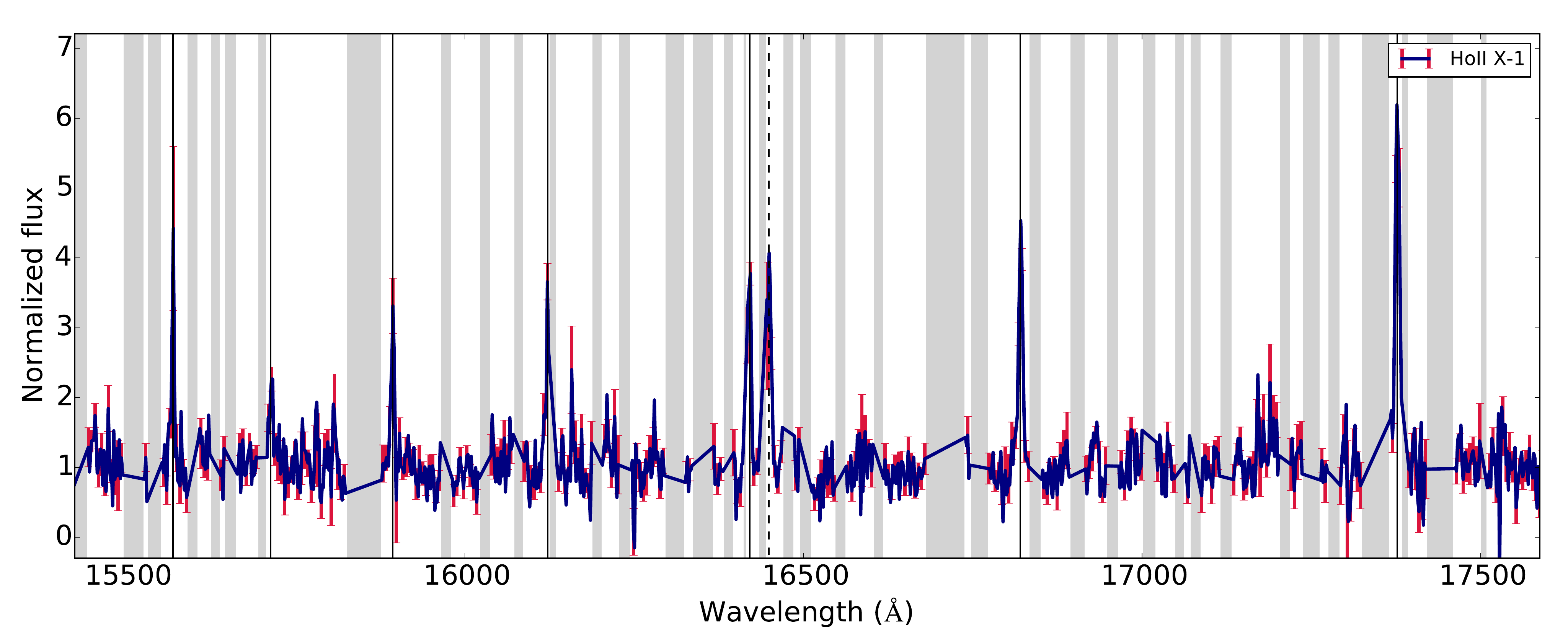}
\caption{The normalized MOSFIRE spectrum of \ho2. The spectrum has been interpolated over wavelength ranges that were strongly affected by noise from telluric emission lines (grey shaded areas). Error bars (in red) are plotted for every third data point. The dashed line indicates the [Fe {\sc ii}] ($\lambda1.644$) line, the solid lines indicate the positions of the hydrogen Brackett lines. The vertical lines representing spectral transitions are redshifted by 165 \vel.}\label{fig:ho2spec}
\end{figure*}

\ho2{} is a well-studied nearby ULX. It is surrounded by a nebula powered by the X-ray source that was discovered in optical observations (e.g.~\citealt{pakull02,lehmann05}). Our NIR spectrum shows a weak continuum with emission lines, most notably the [Fe {\sc ii}] ($\lambda1.644$) line and the hydrogen Brackett lines (see Figure \ref{fig:ho2spec}). We do not significantly detect any absorption lines. The S/N in the continuum is $\sim 4$, which is too low to detect absorption lines from an RSG if they would be present. From the positions of the emission lines we calculate a radial velocity of $165 \pm 3$ \vel, consistent with the radial velocity of the galaxy and earlier observations of the nebula.

\subsection*{NGC 4136: \ulxc}
\begin{figure*}
\includegraphics[width=\textwidth]{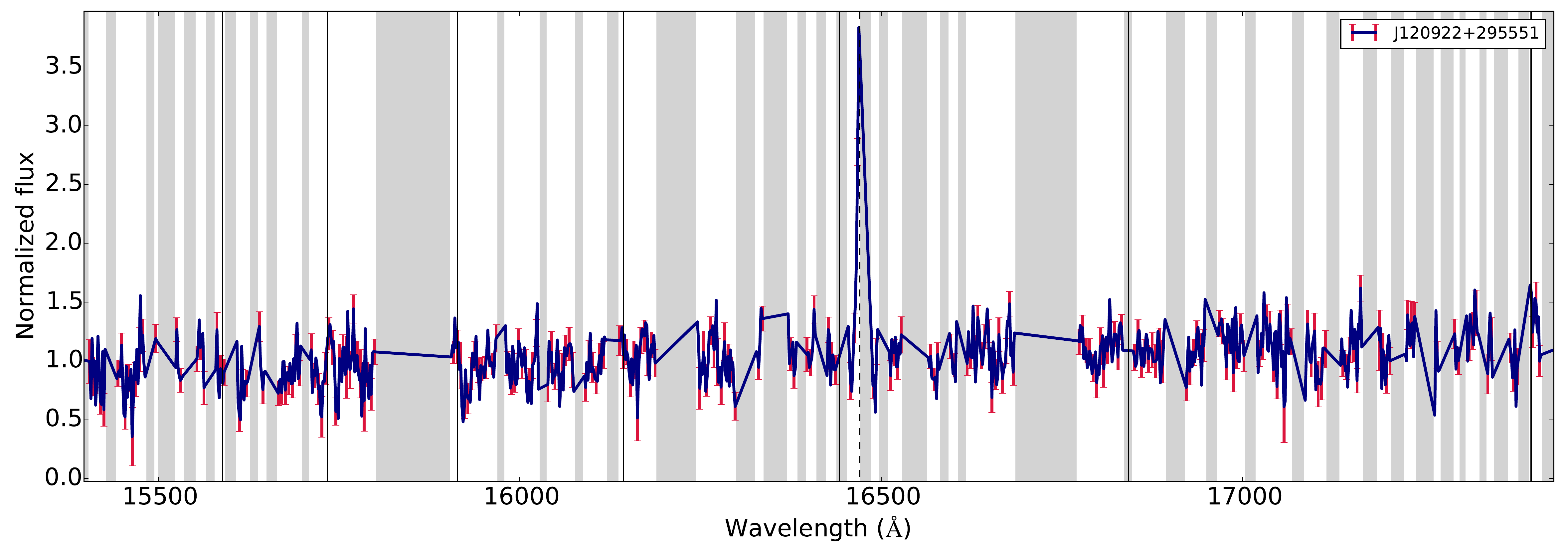}
\caption{Part of the normalized MOSFIRE spectrum of NGC 4136 \ulxc. The spectrum has been interpolated over wavelength ranges that were strongly affected by noise from telluric emission lines (grey shaded areas). Error bars (in red) are plotted for every third data point. The dashed line indicates the [Fe {\sc ii}] ($\lambda1.644$) line, the solid lines indicate the positions of the hydrogen Brackett lines. The vertical lines representing spectral transitions are redshifted by 550 \vel.}\label{fig:ulxcspec}
\end{figure*}

The spectrum of the counterpart of \ulxc{} resembles that of \ho2, with a weak continuum and emission lines (see Figure \ref{fig:ulxcspec}). The 2D spectrum shows spatially extended emission around the position of the ULX, with stronger Brackett emission lines further from the ULX. The radial velocity of the counterpart as measured from the shift of the [Fe {\sc ii}] line is $550 \pm 5$ \vel, proving that the source is located in NGC 4136. The continuum emission could be due to an RSG. The S/N in the continuum is $\sim 5$, and the strongest absorption line in an RSG, the CO-bandhead at 1.62 $\mu$m, lies in a region that is heavily affected by noise from telluric emission lines, precluding us from detecting this feature. Other RSG absorption lines would not be significant at this S/N.

\subsection*{NGC 4136: \ulxd}
\begin{figure*}
\includegraphics[width=\textwidth]{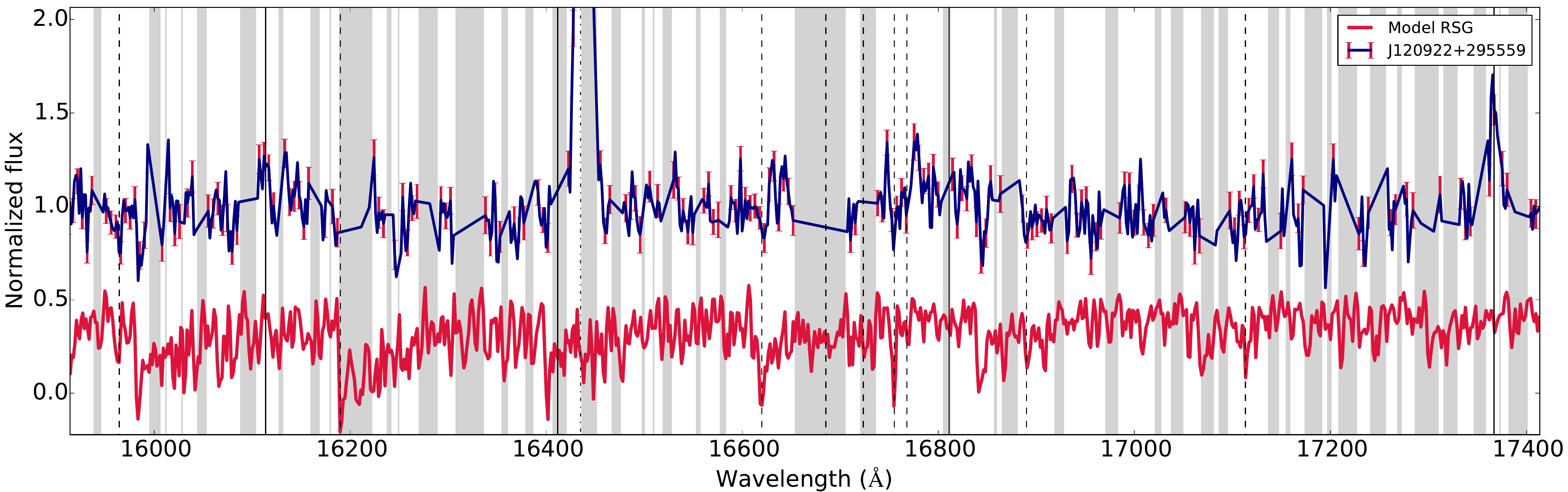}
\caption{The normalized spectra of NGC 4136 \ulxd{} (blue line, with error bars --- in red --- plotted at every third data point), and a 3500 K model from \citet{lancon07} (red line, shifted downwards by 0.5 for clarity). The spectrum of \ulxd{} has been shifted by $-586$ \vel{} and has been interpolated over wavelength ranges that were strongly affected by noise from telluric emission lines (grey shaded areas). Solid lines indicate the positions of hydrogen Brackett lines, dashed lines indicate the positions of absorption lines that are indicative of late-type stars --- from short to long wavelength: Si {\sc i} (1.59 $\mu$m), CO bandhead (1.62 $\mu$m), CO bandhead (1.66 $\mu$m), Si {\sc i} (1.67 $\mu$m), Al {\sc i} triplet (1.672, 1.675, 1.676 $\mu$m), OH (1.69 $\mu$m), and Mg {\sc i} (1.71 $\mu$m). The dash-dotted line indicates the position of the [Fe {\sc ii}] (1.644 $\mu$m) emission line, the peak of which lies at 4.7 (relative units). 
}\label{fig:ngc41362}
\end{figure*}

The spectrum of \ulxd{} shows continuum emission with absorption lines due to neutral metals and CO, indicative of late-type stars (see Figure \ref{fig:ngc41362}). However, the strongest feature in the spectrum is the [Fe {\sc ii}] ($\lambda$1.644) emission line (dash-dotted line), similar to \ulxb{} and \ulxc. Weaker emission lines are visible at the positions of the hydrogen Brackett lines that are in regions uncontaminated by noise from telluric emission lines, most notably at the position of the Br (10-4) line at 1.74 $\mu$m.  
Visual inspection of the 2D spectrum reveals that the peak of the [Fe {\sc ii}] emission is spatially offset by $\sim 0.8''$ with respect to the continuum emission, corresponding to a distance of $\sim 45$ pc. The Br (10-4) line does not show this spatial offset.

In late type stars, the CO bandhead at 1.62 $\mu$m is the strongest absorption feature. Unfortunately, in the \ulxd{} spectrum this bandhead is redshifted into a region heavily contaminated by noise from telluric emission lines, precluding us from reliably detecting it. There is an absorption feature visible at the position of the CO bandhead at 1.66 $\mu$m, as well as at the positions of the Si {\sc i} line at 1.59 $\mu$m, the Al {\sc i} triplet (1.672, 1.675, 1.676 $\mu$m), the OH line at 1.69 $\mu$m, and the Mg {\sc i} line at 1.71 $\mu$m. The cross-correlation with model RSGs yields a radial velocity of $586 \pm 22$ \vel{}, consistent with the radial velocity of NGC~4136 at the location of the ULX (\citealt{fridman05}) and thus proving that the source is indeed located in that galaxy. The highest values for the cross-correlation function are achieved with the templates with temperatures of 2900 -- 4000 K, pointing towards an M-type RSG. We find no evidence for rotational broadening of the absorption lines.

We measure the radial velocity offset of the emission lines by fitting Gaussian profiles to the [Fe {\sc ii}] ($\lambda$1.644) and Br (10-4) lines while masking the parts of the spectrum that contain absorption lines. Fitting the two lines separately we find that their radial velocities are consistent with being the same within the error bars. Fitting them simultaneously to reduce the uncertainty, we find a radial velocity of $570 \pm 4$ \vel{}, consistent with the velocity of the RSG.

%\ulxd{} shows signs of a nebula, possibly powered by the ULX, and absorption features that point towards a late-type stellar companion. These absorption lines may be used in the future to search for radial velocity variations and constrain the mass of the BH. 

% Conclusions
\section{Discussion and conclusions}
We obtained Keck/MOSFIRE \h{} spectra of five ULX counterparts, two located in NGC 925, two in NGC 4136 and \ho2. Two of these, \ulxa{} in NGC 925 and \ulxd{} in NGC 4136 are consistent with being M-type supergiants. The absorption lines in such spectra can in principle be used to measure the radial velocity curve of the counterpart and obtain a lower limit on the BH mass. 

\subsection*{Emission line spectra}
The emission lines in the spectra of \ulxb{}, \ho2{}, \ulxc{} and \ulxd{} can have several origins. They may originate in a nebula, or in the accretion disc --- although the latter is ruled out for the [Fe {\sc ii}] line in \ulxd{}, as it is spatially offset from the spectrum of the counterpart. None of the sources we observed show double-peaked emission lines, which argues against an accretion disc origin for these lines, unless we view the disc under a low inclination angle. 
These emission lines cannot be used for dynamical mass measurements. However, like nebular emission lines observed in optical spectra of some ULXs (cf.~\citealt{pakull02,kaaret04,moon11}) they may be used to infer the true X-ray luminosity of ULXs, and help in determining whether or not the X-ray emission of these sources is beamed. 
The observed line intensity ratio in NGC 925 \ulxb{} of [Fe {\sc ii}] (1.644) to Br-10 is $\sim71.5$. If we use 0.330 as the ratio of Br-10 to Br$\gamma$ line intensities for case B gas in the temperature range of 5000 - 20000 K (\citealt{draine11}), we have $\sim24$ as the ratio of the [Fe {\sc ii}] intensity to Br$\gamma$. This is somewhat lower than what has been observed in several Galactic supernova remnants where the ratio has been estimated to be $\geq27$ and up to $80$ (and possibly higher, see \citealt{koo07} and references therein). Given the relatively low ionization potential, 7.9 eV, of Fe, this indicates that the ionization source of the [Fe {\sc ii}] and HI lines in NGC 925 may be harder than the typical shocks found in Galactic supernova remnants. Strong [Fe {\sc ii}], HI recombination and H$_2$ lines have all been detected in the mid-infrared spectra of the highly-obscured high-mass X-ray binary IGR J16318-4848 (\citealt{moon07}). The lines are believed to be originating in X-ray illuminated dense circumstellar material from the strong mass loss of the optical companion star potentially evolving into an LVB. The similarity supports the interpretation that there is ample material around the ULX.
We will discuss the origin and implications of these emission lines in more detail in a forthcoming paper.

The apparent and absolute magnitudes mentioned in Table \ref{tab:obs} are those of the combined flux in the emission lines and the continuum emission. To estimate the magnitude of a possible donor star, we estimate the magnitude of the continuum. Because we did not observe a flux standard star to calibrate our spectra, we do this by comparing the S/N level for the continuum emission to values calculated with the MOSFIRE exposure time calculator\footnote{http://www2.keck.hawaii.edu/inst/mosfire/etc.html} for our observing conditions and a range of magnitudes. We find that for \ho2, the flux in the emission lines does not contribute significantly to the magnitude of the counterpart. For \ulxb{} in NGC 925, we find a limit on the apparent magnitude of $H \gtrsim 22.5$, corresponding to an absolute magnitude $M_H \gtrsim -7$. This makes it unlikely that the donor star is a supergiant of spectral type F or later (\citealt{tokunaga00}). For \ulxc{} in NGC 4136, we find an apparent magnitude of $H \approx 20.5$, corresponding to an absolute magnitude $M_H \approx -9.4$. If the continuum emission in this source is mainly due to the donor star, this absolute magnitude would mean it is still most likely an RSG. A deeper observation might then reveal stellar absorption lines.

\subsection*{RSG spectra}
The RSG candidate counterpart of the ULX RX~J004722.4-252051 in NGC 253 (\citealt{heida15a}) shows a radial velocity that is offset with respect to its environment by $66 \pm 6$ \vel. The candidate donor stars of \ulxa{} and \ulxd{} do not show such an offset, although source 3 in our \h{} image of the environment of \ulxa{} may show a velocity offset, with an upper limit of $80$ \vel{}, with respect to sources 1 and 2. Analogous to RX~J004722.4-252051 in NGC 253, this offset could be due to binary motion of the RSG around a massive stellar BH or to a natal kick imparted on the BH in the supernova explosion, if source 3 would be the donor star of the ULX. However, this seems unlikely based on the 95\% confidence error circle of the X-ray position of the ULX in Figure \ref{fig:ulxa_slits}. Alternatively, source 3 may be a runaway RSG (e.g.~\citealt{eldridge11}).

Among these three candidate donor stars, \ulxd{} in NGC 4136 is the only one that also shows emission lines in its NIR spectrum, strengthening its association with the ULX. Optical emission lines from an H {\sc II} region close to the ULX are visible in the spectrum of RX~J004722.4-252051 in NGC 253, but no emission lines were detected in the NIR part of its spectrum. The counterpart to \ulxa{} in NGC 925 also does not show NIR emission lines. Following the same procedure as in \citet{heida15a}, we calculate the probability that this NIR source is an interloper that happens to coincide with the ULX. We compute the density of NIR sources that are as bright or brighter than the ULX candidate counterpart in our WHT/LIRIS \h{} image (see \citealt{heida14} for details). We only consider the bottom 1/3 of the image, because NGC 925 occupies that part of the detector. In this area of $2 \times 10^ 4$ arcsec$^2$ we detect 99 sources. The $95\%$ confidence error circle of the position of the X-ray source on the \h{} image has an area of $4.1$ arcsec$^2$. Thus the probability of a chance superposition is approximately $2\%$. Based on this we cannot exclude that the NIR source is an interloper, although the probability is rather low.

The counterpart of \ulxa{} is elongated in our \h{} image, but during our Keck/MOSFIRE observations, seeing conditions were not good enough to allow us to separate the spectra of the two putative components. The peak-to-peak distance of the elongated core in our LIRIS image is $\sim0.7''$. An adaptive optics-assisted or {\it Hubble Space Telescope} NIR image at $\sim0.1''$ resolution would be useful to confirm whether or not the counterpart in our LIRIS image consists of two individual sources. If this is the case, a bore-sight corrected \chan{} position of the ULX, with a localization error of $\sim0.2''$, could distinguish whether one of these is the counterpart of the ULX or if the NIR source and ULX are unrelated.

\begin{figure*}
\hbox{
\includegraphics[height=5.5cm]{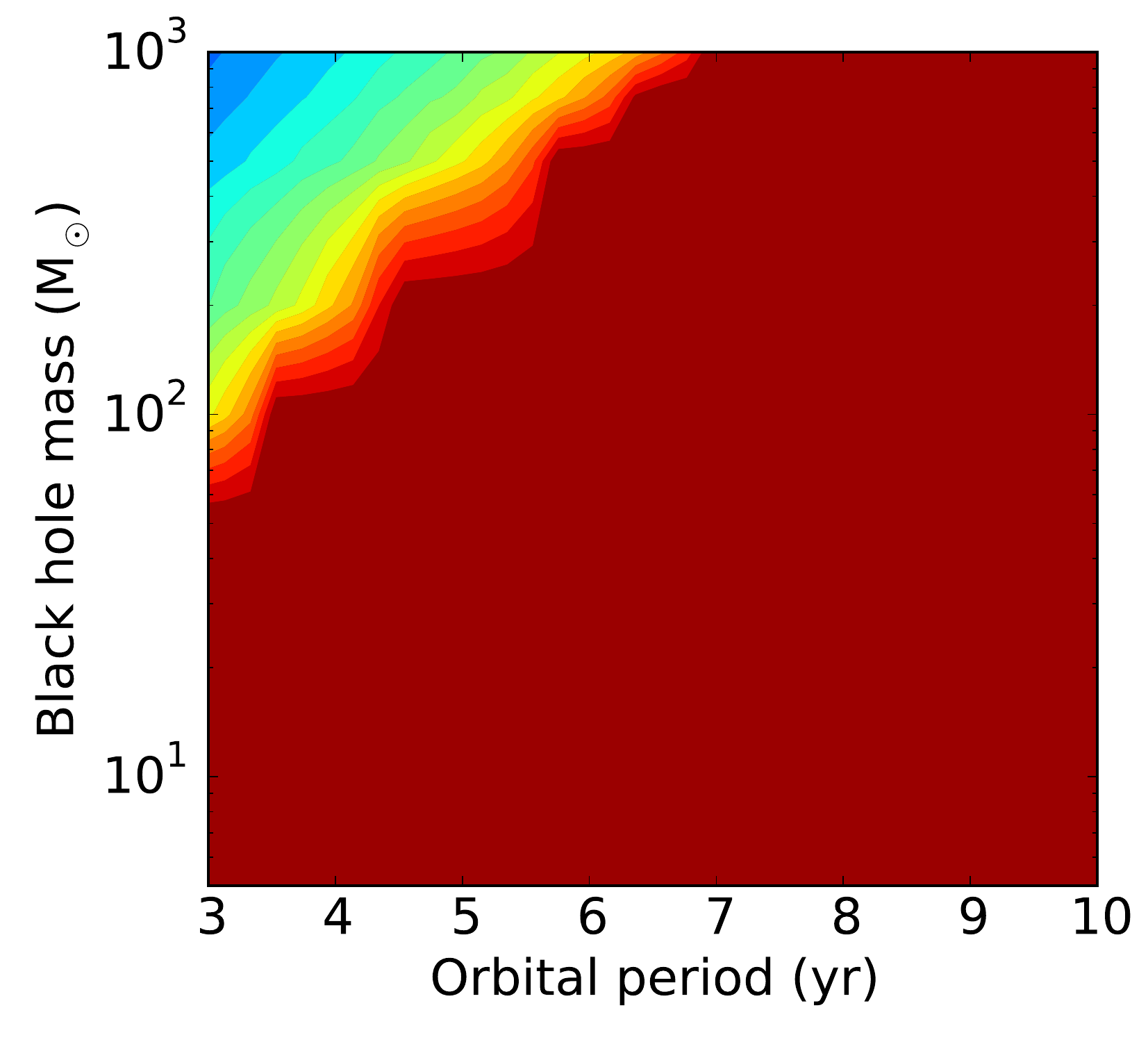}
\includegraphics[height=5.5cm]{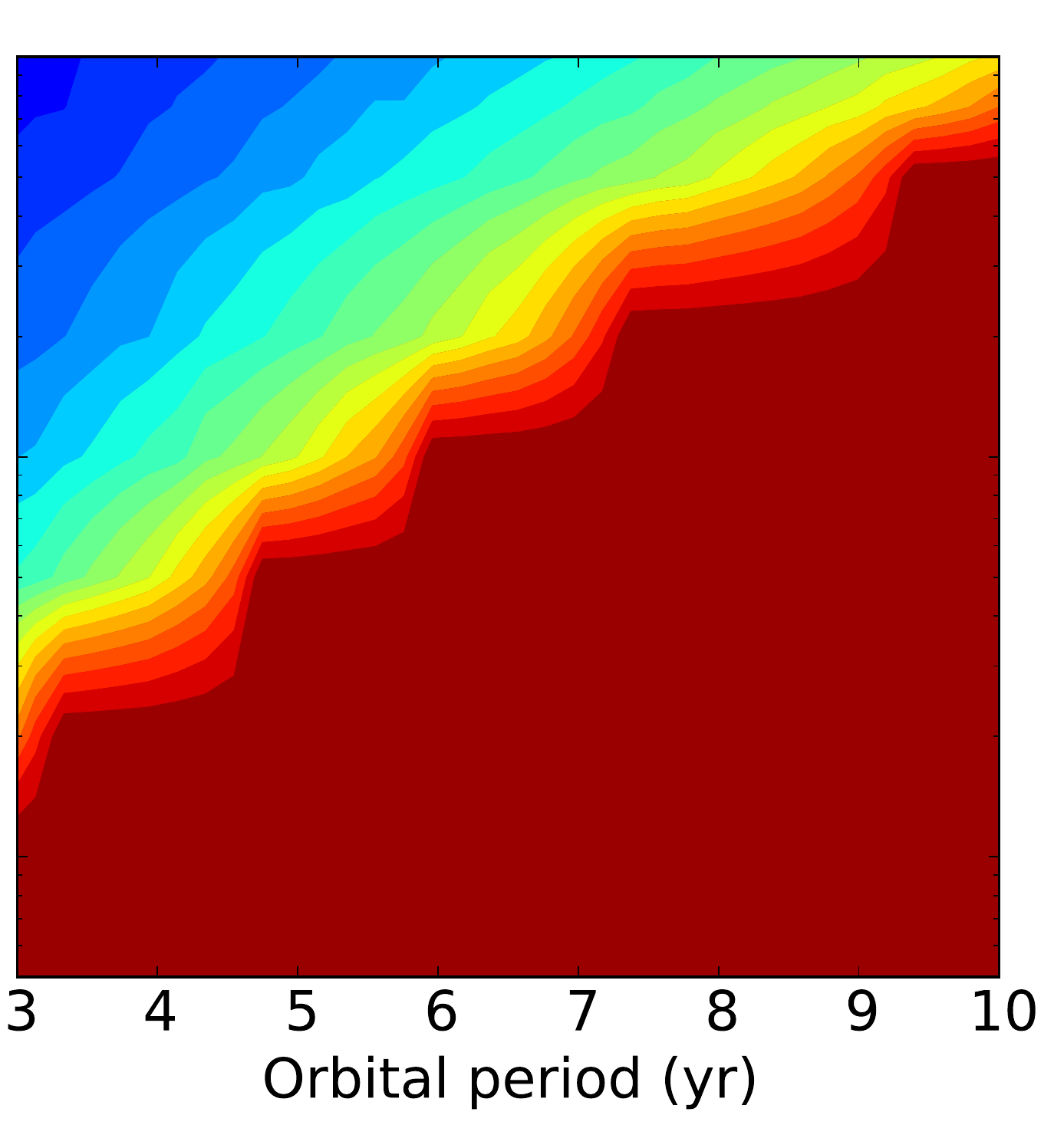}
\includegraphics[height=5.5cm]{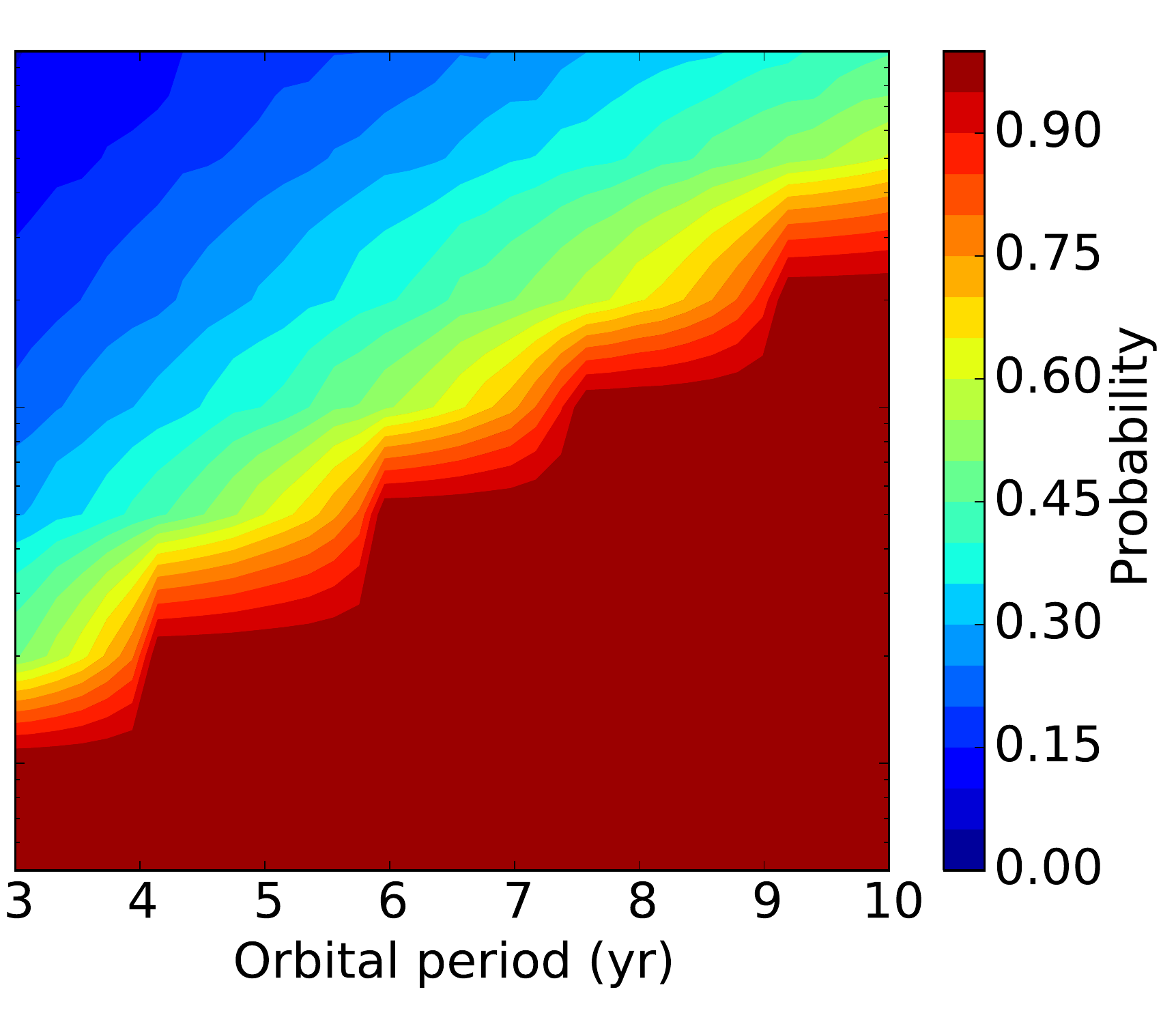}
}
\caption{The probability that the radial velocities measured in our two spectra of NGC 925 \ulxa{} obtained 10 months apart are consistent within 40 \vel, as a function of BH mass and orbital period, for system inclinations of 20$^\circ$, 40$^\circ$ and 60$^\circ$ (left to right). We assume a mass of 10 \msun{} for the RSG and that mass transfer happens through Roche lobe overflow. Orbital periods shorter than 3 years are ruled out given the radius of an M-type RSG. The step-like shape is due to the discrete sampling of BH masses.}\label{fig:mc-velo}
\end{figure*}

The ultimate proof that the RSG is orbiting a black hole can only come from the detection of periodic shifts of its radial velocity. We obtained two epochs of spectroscopy of the candidate counterpart and found no change in the radial velocity, with a 2-$\sigma$ upper limit of 40 \vel. Assuming that the RSG is the donor star in a binary system, the probability that we would not detect a velocity shift between two measurements depends on the orbital period, ratio of the mass of the donor star to the mass of the BH, and the inclination of the system. We calculate this probability by drawing 10000 random values for the orbital phase at the time of the first observation, and calculating how many of these result in a velocity shift smaller than our observed upper limit for a range of BH masses, orbital periods and system inclinations.
Here we assume that mass transfer happens through Roche lobe overflow, as wind-feeding is unlikely to be a viable mechanism for ULXs (\citealt{copperwheat07}).The results of this calculation are shown in Figure \ref{fig:mc-velo} for system inclinations of 20$^\circ$, 40$^\circ$ and 60$^\circ$. The upper limit of 40 \vel, with the measurements taken 308 days apart, puts very weak restraints on the orbital period or BH mass if the system inclination is $< 40^\circ$. For $i = 40^\circ$, BH masses larger than $\sim100$ \msun{} in combination with orbital periods shorter than $\sim4$ years would likely (probability $> 50\%$) have led to a velocity difference larger than 40 \vel{} and would thus have been detected. For $i \geq 60^\circ$, the same BH masses in combination with orbital periods shorter than $\sim6$ years would likely have led to a detection. Given the expected radius of an M-type RSG, orbital periods shorter than 3 years can be ruled out as the Roche-radius would then be too small to contain the RSG.
If the RSG is the donor star of the ULX, the most likely options are that either the system is seen at low inclination, or the BH mass is less than 100 \msun, unless the orbital period is longer than 6 years.

The stellar absorption lines in the spectra of \ulxa{} and \ulxd{} make them excellent targets for dynamical mass measurements of black holes in ULXs, if these RSGs are indeed the donor stars of the ULXs. However, as the non-detection of a radial velocity shift in \ulxa{} shows, we will need observations over time spans of many years to cover the expected orbital periods of these systems: first to confirm or rule out that these RSGs show radial velocity variations, and if they do, to measure their full radial velocity curve and set a lower limit to the mass of the BH.

\section*{Acknowledgements} 
MH would like to thank Nick Konidaris for his help with the MOSFIRE DRP. We thank Tom Marsh for developing {\sc Molly}. TPR acknowledges support from STFC as part of the consolidated grant award ST/L00075X/1. The work of DJW and DS was carried out at Jet Propulsion Laboratory, California Institute of Technology, under a contract with NASA. PGJ acknowledges support from ERC consolidator grant number 647208. The data presented herein were obtained at the W.M. Keck Observatory, which is operated as a scientific partnership among the California Institute of Technology, the University of California and the National Aeronautics and Space Administration. The Observatory was made possible by the generous financial support of the W.M. Keck Foundation. The authors wish to recognize and acknowledge the very significant cultural role and reverence that the summit of Mauna Kea has always had within the indigenous Hawaiian community.  We are most fortunate to have the opportunity to conduct observations from this mountain.

% The bibliography
\bibliographystyle{mnras}
\bibliography{bibliography}

% Don't change these lines
\bsp	% typesetting comment
\label{lastpage}
\end{document}